\documentclass[]{aa}
\usepackage{graphicx}
\usepackage{txfonts}
\begin{document}
   \title{Tracing the spiral arms in IP~Pegasi}

  \author{R. Baptista\inst{1},
          L. Morales-Rueda\inst{2},
          E.~T. Harlaftis\inst{3}\fnmsep
	  \thanks{in memoriam},
          T.~R. Marsh\inst{4}
          \and
	  D. Steeghs\inst{5}
          }

   \offprints{R. Baptista, e-mail: bap@astro.ufsc.br}

    \institute{Departamento de F\'{\i}sica, Universidade Federal de Santa
     Catarina, Campus Trindade, 88040-900, Florian\'opolis, SC, Brazil\\
     \email{bap@astro.ufsc.br}
    \and
     Department of Astrophysics, Radboud University Nijmegen, 6500 GL 
     Nijmegen, The Netherlands\\
     \email{lmr@astro.kun.nl}
    \and
     Institute of Space Applications and Remote Sensing, National 
     Observatory of Athens, P. O. Box 20048, Athens 118 10, Greece\
    \and
     Department of Physics, University of Warwick, Coventry CV4 7AL, UK\\
     \email{T.R.Marsh@warwick.ac.uk}
    \and
     High Energy Astrophysics Division, Center for Astrophysics, MS-67, 
     60 Garden Street, Cambridge, MA 02138, USA\\
     \email{dsteeghs@head.cfa.harvard.edu}
}

   \date{Received 2005 May 11; accepted 2005 July 1}

   \abstract{

We report the analysis of time-resolved spectroscopy of IP~Pegasi in 
outburst with eclipse mapping techniques to investigate the location
and geometry of the observed spiral structures.  We were able to obtain
an improved view of the spiral structures with the aid of light curves 
extracted in velocity bins matching the observed range of velocities 
of the spiral arms combined with a double default map tailored for 
reconstruction of asymmetric structures.
Two-armed spiral structures are clearly seen in all eclipse maps.
The arms are located at different distances from the disc centre.
The ``blue'' arm is farther out in the disc ($R= 0.55 \pm 0.05\;R_{L1}$) 
than the ``red'' arm ($R= 0.30 \pm 0.05\;R_{L1}$).
There are evidences that the velocity of the emitting gas along the
spiral pattern is lower than the Keplerian velocity for the same disc 
radius.  The discrepancy is smaller in the outer arm (measured velocities
10-15 per cent lower than Keplerian) and is more significant in the 
inner arm (observed velocities up to 40 per cent lower than Keplerian).
We measured the opening angle of the spirals from the azimuthal 
intensity distribution of the eclipse maps to be $\phi= 25^o\pm 3^o$.
A comparison with similar measurements on data at different outburst
stages reveals that the opening angle of the spiral arms in IP~Peg 
decreases while the outbursting accretion disc cools and shrinks, 
in agreement with the expected evolution of a tidally driven spiral wave.
The sub-Keplerian velocities along the spiral pattern and the clear
correlation between the opening angle of the spirals and the outburst
stage favors the interpretation of these asymmetric structures
as tidally-induced spiral shocks.

 }

\titlerunning{Tracing the spiral arms in IP Pegasi}
\authorrunning{Baptista et al.}

   \maketitle
%

\section{Introduction}

IP Pegasi is a dwarf nova, a compact binary in which a late-type star 
overfills its Roche lobe and transfers matter to a companion white dwarf 
via an accretion disc.  It shows recurrent outbursts (every 70-100 days) 
in which the disc expands and brightens by 2-3 magnitudes during 10-12 
days as a consequence of a sudden increase in mass accretion through 
the disc.  In the quiescent, low-mass accretion state, the white 
dwarf and the bright spot (formed by the impact of the infalling gas 
stream with the outer edge of the disc) dominate the light from the 
binary at optical and ultraviolet wavelengths, whereas during outburst 
the hot, large disc is the dominant source of light.  
The binary is seen at a high inclination angle ($i=81^o$), which 
leads to deep eclipses in the light curve every 3.8 hours when the 
white dwarf, accretion disc and bright spot are occulted by the mass-donor 
star.  This allows the emission from the different light sources to be 
distinguished and spatially resolved studies to be performed,
making IP~Peg an ideal laboratory for the study of accretion physics.
In particular, it is well suited for the application of indirect imaging
techniques to resolve the disc emission both in position (eclipse mapping, 
Horne 1985) and velocity (Doppler tomography, Marsh \& Horne 1988).

Additional attention has been drawn to IP~Peg after the discovery of a
two-armed spiral structure in its accretion disc during outburst (Steeghs 
et~al. 1997).  These spiral structures were seen in different outbursts 
and at different outburst stages -- from the end of the rise and maximum 
(Steeghs et~al. 1997, Harlaftis et~al. 1999, Baptista et~al. 2000) 
through the decline stages (Morales-Rueda et~al. 2000, Baptista et~al. 
2002) and possible also at the very late decline (Saito et~al. 2005).  
They are believed to be raised by tides from the mass-donor 
star on the outer accretion disc when it expands during outburst
(see review by Steeghs 2001).

Spiral shocks in accretion discs were predicted theoretically and 
advocated as an alternative mechanism for the removal of angular 
momentum required for the disc gas to accrete (Sawada et~al. 1986).  
Numerical hydrodynamic simulations suggest that strong spiral shocks 
are formed in hot accretion discs (e.g., Sato et~al. 2003 and references
therein) and that the opening angle of the spirals (the angle formed by 
the shock front and the azimuthal direction of the disc gas flow) 
depends on the disc temperature: the cooler the disc, the more tightly 
wound the spiral pattern is.  Godon et~al. (1998) argued that wide open 
spirals could only form in accretion discs much hotter ($T\sim 
10^6\;K$) than those of dwarf novae.  Steeghs \& Stehle (1999) remarked 
that the disc radius is also important in determining the shape of the
spirals and that wide open spiral arms such as those observed in
IP~Peg could be raised at the much lower temperatures found in
accretion discs of outbursting dwarf novae ($T\simeq 10^4\;K$) if 
the disc is large -- as expected for outbursting discs.

In a previous paper (Harlaftis et~al. 2004) we investigated the
ability of the eclipse mapping method to recover spiral structures and 
we set the observational requirements for a successful reconstruction. 
We also showed how the detection of spiral arms with eclipse mapping
is improved with a proper extraction of velocity-resolved eclipse
light curves to avoid contamination by light sources other than the
spiral arms.  This is the follow up paper.
Here we reanalyze the spectroscopic data of Morales-Rueda et~al. (2000)
with eclipse mapping techniques to investigate the spatial properties
of the observed two-armed spiral structure.  

This paper is organized as follows.  The details of the analysis
are presented in section~\ref{analise} while the results are presented
and discussed in section~\ref{results}.  Section~\ref{discuss} compares
the orientation of the spiral arms from this study with those from the
previous analysis of Baptista et~al. (2000, hereafter BHS) and Baptista
et~al. (2002, hereafter BHT) to infer the changes in opening angle of 
the spiral arms with outburst stage, and discusses the implications
of our results to the interpretation of the observed spiral structures 
in terms of irradiation of tidally thickened outer disc regions by 
a hot inner disc. 
The conclusions are summarized in section~\ref{conclusions}.  
The Appendix discusses the changes implemented in the eclipse mapping 
method in order to allow an improved reconstruction of asymmetric 
structures such as spiral arms in accretion discs.

\section{Data analysis} \label{analise}

\subsection{Observations}
\label{observa}

Time-resolved spectro-photometry of IP~Pegasi was secured with the
Intermediate Dispersion Spectrograph on the 2.5\,m Isaac Newton
Telescope, at La Palma, during the nights of 1994 October 30 and 31.
The observations were made 5 and 6 days after the onset of an
outburst, when the star was at the maximum brightness plateau phase 
of the 12-14 days long outburst (Morales-Rueda et~al. 2000, see 
their Fig.\,1).  A total of 164 blue spectra ($\lambda\lambda\; 
4040 - 4983 \AA$) was obtained at a time resolution of 220\,s and a 
spectral resolution of 100 km s$^{-1}$ at H$\beta$.  The reader is 
referred to Morales-Rueda et~al. (2000) for a detailed description of 
the data set and of the reduction procedures.

The spectra were phase-folded according to the linear ephemeris
of Wolf et~al. (1993),
\begin{equation}
T_{mid}(HJD) = 2\,445\,615.4156 + 0.15820616\,E
\label{efem}
\end{equation}
where $T_{mid}$ is the inferior conjunction of the secondary star.
Our observations are bracketed by two sets of HST observations of
IP~Peg in quiescence (Baptista et~al. 1994) in which the white dwarf 
eclipse egress times can be precisely measured.  From these timings,
we infer that the white dwarf mid-eclipse at that epoch occurred
0.008 orbital cycles before the prediction of the ephemeris in
Eq.(\ref{efem}) (assuming a white dwarf eclipse width of $\Delta\phi= 
0.086$ cycles, Wood \& Crawford 1986).  Therefore, an offset of $+0.008$ 
cycles was added to the phases in the light curve to make phase zero 
coincident with the inferior conjunction of the secondary star.  The 
observations cover the eclipse cycles $E=25158, 25159, 25164$ and 25165.

\subsection{Emission-line light curves and eclipse mapping}
\label{cluz}

\begin{figure}
\centering
\includegraphics[bb=3cm 5cm 19cm 21.5cm,angle=-90,scale=0.42]{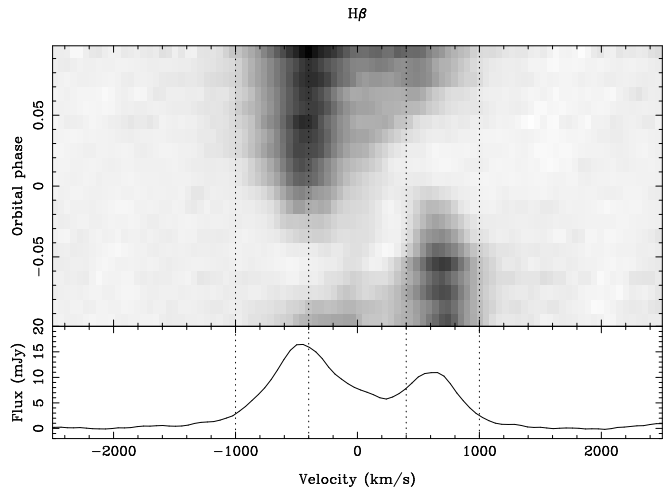}
\caption{ Top: zoomed trailed spectrogram of the H$\beta$ line around 
   eclipse.  Bright regions are dark.  Bottom: the integrated line profile 
   along the eclipse phases of the spectrogram on the top panel (from 
   $\phi= -0.1$ to $+0.1$~cycle).  Vertical dotted lines mark the two
   velocity-resolved regions used to extract light curves of the spiral 
   arms and to isolate low-velocity emitting gas. }
\label{fig1}
\end{figure}

The upper panel of Fig.~\ref{fig1} shows a zoomed trailed spectrogram 
of $H\beta$ during eclipse (phase range $-0.1$ to $+0.1$ cycles) on 
the first night of observations. The lower panel shows the resulting 
line profile integrated over the eclipse phase range.  
The emission lines of IP~Peg in outburst have contributions from the 
spiral arms as well as the irradiated face of the secondary star 
(Harlaftis 1999) and possibly another low-velocity emission component 
(Harlaftis et~al. 1999, Steeghs et~al. 1996).  The signature of the 
two-armed spiral structure appears in Fig.~\ref{fig1}
as the two bright arcs running from positive, high velocities 
($v\geq 400\; km\,s^{-1}$) before mid-eclipse to negative, high 
velocities after phase zero.  The observed change in line profile during 
eclipse indicates that the ``red'' spiral arm {\bf is} being eclipsed at 
about the same phase ($\phi \sim -0.01$ cycles) at which the ``blue'' 
arm starts reappearing from eclipse.  The low-velocity emission is 
seen as the faint line-emission component changing from blue to red shifts 
(at $v \leq 400\;km\,s^{-1}$) around mid-eclipse.
%
%

Because the spiral arms and the other emission components contribute to 
the line profile at different Doppler velocities, it is possible to 
separate the contribution of these distinct light sources by a proper 
extraction of velocity-resolved eclipse light curves.  
For example, BHT performed velocity-resolved eclipse mapping to 
separate the HeI emission from each of the spiral arms in IP~Peg.
Harlaftis et~al. (2004) discussed how dilution by other light sources
reduces the ability of the eclipse mapping method to reconstruct spiral
structures in accretion discs.  They also showed how the detection of 
spiral arms may be improved with the aid of velocity-resolved light 
curves.  Here we use this technique as an useful tool to minimize the 
contribution of the low-velocity gas to the light curve and to 
optimize the detection of the spiral arms in line eclipse maps. 

Fig.~\ref{fig1} shows the choice of the velocity-resolved passbands 
used to separate the contribution of the spirals from that of 
low-velocity gas in the emission lines of IP~Peg.
Two light curves were extracted for each line, one covering the low
velocity range ($-400$ to $+400\;km\,s^{-1}$) and one matching the 
observed velocity range of the spiral arms ($-1000$ to $-400$ plus 
$+400$ to $+1000\; km\,s^{-1}$ for all lines except He\,I 4472, for
which we use the narrower ranges $-900$ to $-400$ plus $+400$ to 
$+900\; km\,s^{-1}$ to avoid contamination by the Mg\,II $\lambda 
4481$\AA\ line).  The low-velocity range light curve will be
hereafter referred to as the 'core' light curve, while the higher
velocity range curve will be referred to as the 'spiral' light curve.
Core and spiral light curves were extracted for H$\beta$, H$\gamma$, 
He\,II 4686, He\,I 4472 and the Bowen blend at $\lambda\lambda\;
4640-4660$ \AA.

\begin{figure*}
\includegraphics[bb=0.5cm 0cm 19.5cm 18cm,angle=-90,scale=0.6]{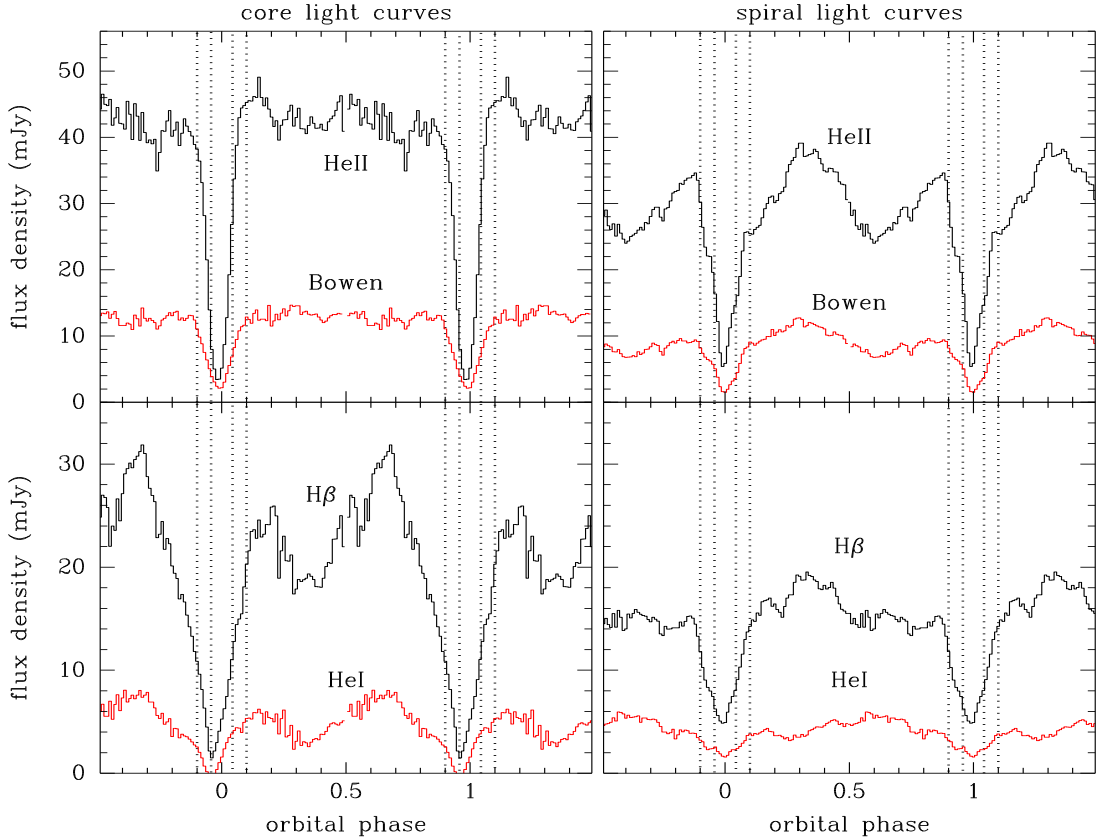}
\caption{ Extracted core (left panels) and spiral (right panels) light 
  curves for He\,II and H$\beta$ (in black) and for He\,I and the Bowen 
  blend (in grey).  Vertical dotted lines mark mid-ingress/egress phases 
  of the white dwarf and the reference phases $-0.1$ and $+0.1$ cycles.  
  The light curves are repeated in phase for visualization purposes. 
  The H$\gamma$ light curves (not shown) are very similar to those of 
  H$\beta$.}
\label{fig2}
\end{figure*}

Four orbital cycles were observed.  The orbital modulation and the
shape of the eclipse are consistent from one night to the other
(see Fig.~5 of Morales-Rueda et~al. 2000), indicating that the disc 
brightness distribution is relatively stable over that time scale.  
We therefore combined the 4 light curves to produce median light curves 
at a phase resolution of 0.01 cycles for all lines.  The error bars were 
computed from the absolute deviation with respect to the median in 
each phase bin. The resulting light curves are shown in Fig.~\ref{fig2}.
%
%

The He\,II and Bowen blend core light curves show a narrow, deep and
reasonably well centered eclipse indicating that the low-velocity 
emission in these lines arise from a region close to the white dwarf
at disc centre.  A possible interpretation would be that this 
line component is produced in a collimated, hot wind from the inner 
disc, in which most of the velocity is directed away from the orbital 
plane giving rise to small Doppler shifts.  
The Balmer and He\,I core light curves show the very broad, asymmetric 
and phase offset eclipse seen in Harlaftis et~al. (2004, see their 
Figs.\,7 and 8).  The eclipse is too wide to be caused by occultation 
of the accretion disc by the secondary star.  The shape of this eclipse 
is possibly produced by self-occultation
of an asymmetric brightness distribution in the irradiated face of
the secondary star or by occultation of an emission source at a 
different position in the binary (e.g., Steeghs et~al. 1996).
On the other hand, all spiral light curves show the orbital modulation 
and asymmetric eclipse shape characteristics of the emission from a 
two-armed spiral structure in the accretion disc (e.g., BHS, Steeghs 
\& Stehle 1999).  Two separate brightness sources are eclipsed in a 
sequence.  The first one goes in eclipse at $\phi_{1i} \simeq -0.09$~cycle 
and the second one is occulted at $\phi_{2i} \simeq -0.02$~cycle; 
they reappear from eclipse at phases $\phi_{1e}\simeq +0.02$~cycle and 
$\phi_{2e}\simeq +0.07$~cycle, respectively (see Fig.~\ref{fig3} for 
a zoomed view of the light curves around eclipse).  
Hence, our choice of passband successfully framed the emission from 
the spiral arms in the spiral light curves
\footnote{Similar spiral light curves and eclipse shapes are obtained
if we extend the spiral passband to higher velocity ranges (up to 
$|v| = 1300\;km\;s^{-1}$, see Harlaftis et~al. 2004).  However, 
the resulting light curves have lower signal-to-noise ratios as a 
consequence of the reduced average flux. }.

Morales-Rueda et~al. (2000) found that the eclipses of the emission
lines were significantly shifted towards negative phases. The +0.008~cycle
offset applied to the data (section~\ref{observa}) eliminates the 
phase offset of the continuum light curve and reduces the discrepancy 
for the line light curves.  Fig.\ref{fig2} shows that the remaining 
phase offset is caused by asymmetries in the brightness distribution 
of the low-velocity line emitting gas (core light curves). There is no
phase offset in the spiral light curves.  Therefore, the shifted eclipses
seen in the full-line light curves are not an evidence of unequal 
brightnesses in the two arms, as claimed by Smak (2001).

In the remainder of this paper we concentrate on the analysis of the
spiral light curves.  
Our eclipse mapping code (Baptista \& Steiner 1993) assumes that any 
brightness change in the light curve is caused by occultation of disc 
regions by the secondary star.  Therefore, the out-of-eclipse orbital 
modulation of the spiral light curves was removed by fitting a spline 
function to the regions of the light curve excluding the eclipse, 
dividing the light curve by the fitted spline, and scaling the result 
to the flux level of the spline function at phase zero (e.g., BHS).  
This procedure allows the removal of orbital modulations
with only minor effects of the eclipse shape itself.

The eclipse maps are flat square arrays of $51 \times 51$ pixels and 
side $2\, R_{L1}$ centered on the white dwarf, where $R_{L1}$ is the
distance from disc centre to the inner Lagrangian point.  The adopted 
eclipse geometry is $i=81^o$ and $q=0.5$ (Wood \& Crawford 1986),
which ensures that the white dwarf is at the centre of the map.  
The reconstructions were obtained using a double default function 
with $\Delta\theta=10^o$ (See Appendix).

The statistical uncertainties of the eclipse maps were computed with a 
Monte Carlo procedure (e.g., Rutten et~al. 1992).  For each line light 
curve, we create a set of 20 artificial light curves in which the flux 
at each phase is varied around the true value according to a Gaussian 
distribution with standard deviation equals to the uncertainty at that 
point.  These light curves are fitted with the eclipse mapping code to 
generate a set of randomized eclipse maps.  These are combined to produce 
an average map and a map of the standard deviations with respect to the
average.  A map of the statistical significance (or the inverse of the
relative error) is obtained by dividing the true eclipse map by the
map of the standard deviations.

\section{Results} \label{results}

\begin{figure*}
\includegraphics[bb=0cm 3cm 16cm 23.5cm,scale=0.85]{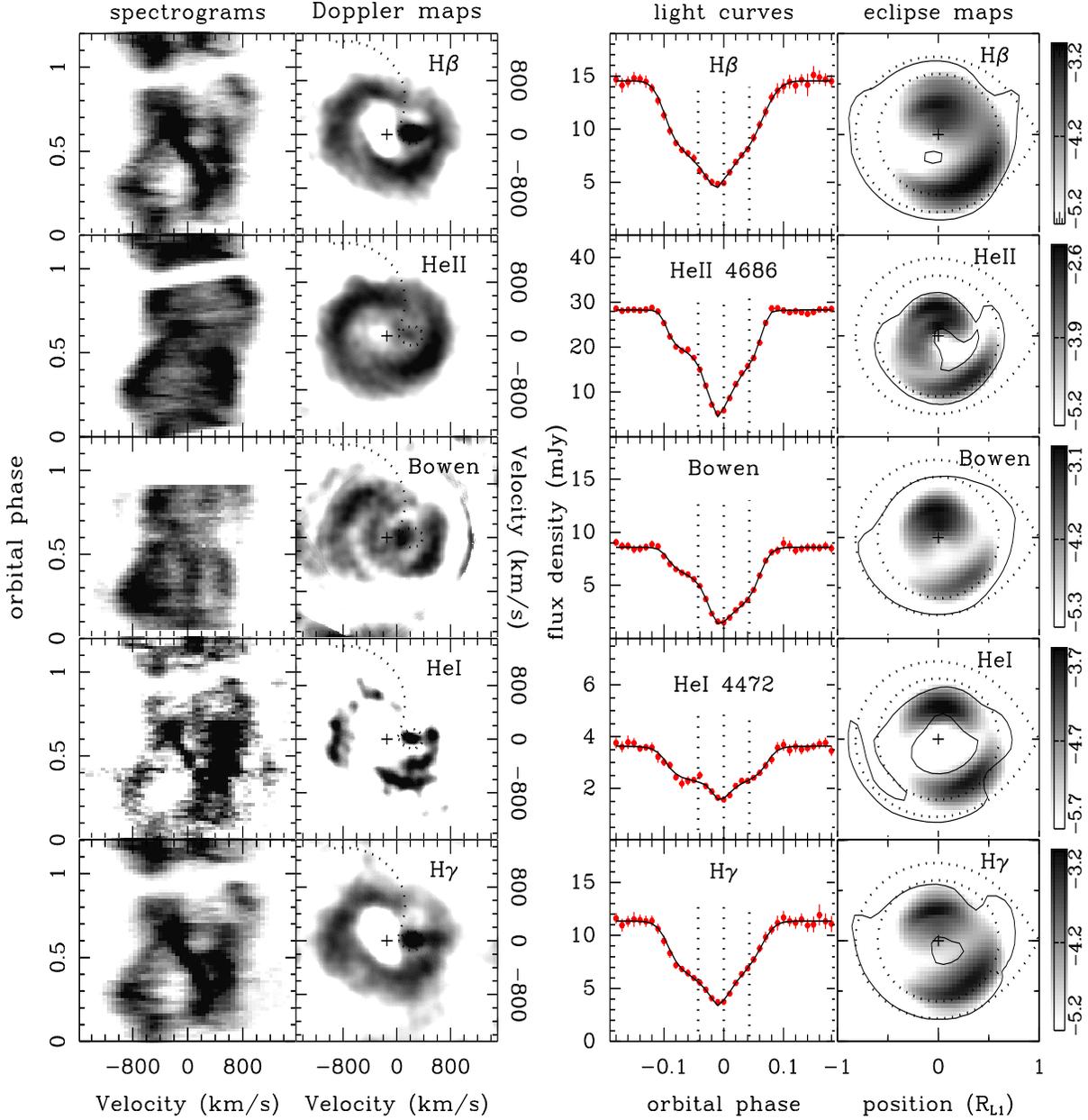}
\caption{  Left: Data trailed spectrograms (linear greyscale) and 
   corresponding Doppler maps (logarithmic greyscale) for H$\beta$, 
   He\,II 4686, the Bowen blend, He\,I 4472, and H$\gamma$.
   Dark regions are brighter.  A cross marks the position of the white
   dwarf.  Dotted lines outline the Roche lobe of the secondary star 
   and the velocities along the ballistic stream trajectory.  Right: Line 
   spiral light curves and corresponding eclipse maps on a logarithmic 
   greyscale.  Dark regions are brighter.  The bar in the right side 
   indicates the greyscale used in each case.  Dotted lines show the 
   primary Roche lobe and a circle of radius $0.6 \;R_{L1}$.  A cross
   marks the centre of the disc.  The secondary is to the right of each
   panel and the stars rotate counter-clockwise.  A solid contour line 
   is overploted on each eclipse map to indicate the 3-$\sigma$ 
   confidence level region. }
 \label{fig3}
\end{figure*}

The resulting eclipse maps are shown in Fig.~\ref{fig3} in a logarithmic
grey-scale.  For completeness, the data trailed spectrograms and the
computed Doppler maps of the corresponding lines (Morales-Rueda et~al.
2000) are shown in the left two columns.  Doppler maps for the $H\gamma$ 
and Bowen blend lines (not shown in Morales-Rueda et~al. 2000) are also 
included. The Doppler maps and trailed spectrograms correspond to the 
data of the first night of observations, except for the Bowen blend map.
In this case the data of the two nights were combined and the spectra
around eclipse excluded from the analysis in order to produce a single 
Doppler map of better signal-to-noise ratio.
%
%
The Doppler maps of Fig.~\ref{fig3} are shown rotated 90 degrees 
clockwise with respect to their usual presentation (e.g., Morales-Rueda
et~al. 2000) in order to display the spiral arms in a similar 
orientation as in the eclipse maps.  The blue spiral arm appears 
closer to the secondary star in the lower right side of the Doppler map.
We shall remark that the coordinate frames of the Doppler and eclipse 
maps are very different: the inner, low-velocity side of a spiral
arm in the Doppler map is related to the outer side of the 
corresponding structure in the eclipse map.

The advantage of using velocity-resolved light curves matched to 
the observed velocity range of the spirals becomes clear when we
compare our eclipse maps with those obtained from light curves over 
the full line width (BHS; Harlaftis et~al. 2004).
BHS needed to compute the asymmetric part of their eclipse maps in 
order to see clearly the spiral structures, whereas the significant 
dilution by other line emitting sources hampered the detection of the 
spiral arms in the Balmer and He\,I full-line eclipse maps of 
Harlaftis et~al. (2004).  Here, the spiral structures are conspicuous 
in the eclipse maps without the need of further data processing.

The double-stepped bulges in the eclipse shapes map into two asymmetric 
structures corresponding to the two-armed spiral structure seen in the 
Doppler maps.  The solid contour line overploted on each eclipse map 
depicts the 3-$\sigma$ confidence level region as derived from the map 
of the statistical significance in each case (section~\ref{cluz}). The 
two-armed spiral structure is at or above the 3-$\sigma$ confidence 
level in all line maps.  We will hereafter refer to these structures 
as to the ``blue'' and ``red'' spiral arms.  
They are located at different distances from disc centre. 
The red spiral arm is seen in the upper side of the eclipse map, at 
radii of $0.2-0.4\;R_{L1}$.  The blue arm is seen in the lower right 
side of the map, at radii of $0.45-0.65 \;R_{L1}$.  This is in
agreement with the Doppler tomography, which indicates smaller
velocities for the blue spiral arm ($\simeq 600\;km\, s^{-1}$) than 
for the red spiral arm ($\sim 700\;km\, s^{-1}$) [Morales-Rueda
et~al. 2000].  The red arm is brighter than the blue arm in the Bowen 
blend, whereas the opposite holds for $H\beta$.  The two arms have 
comparable brightness in the other lines.  This is also in line with 
the results from the Doppler tomography.

To further investigate the properties of the asymmetric arcs we
divided the eclipse map into azimuthal slices (i.e., `slices of pizza')
and computed the radius at which the intensity is a maximum for each
azimuth.  A corresponding Keplerian velocity is obtained from the 
radius of maximum intensity assuming that $M_1= 1.02\;M_\odot$ and 
$R_{L1}= 0.81\;R_\odot$ for IP~Peg (Marsh \& Horne 1990).
This exercise allows us to trace the distribution of the spiral
structures in radius and azimuth (BHS).

\begin{figure*}
\includegraphics[bb=0.5cm 0cm 18.5cm 18cm,angle=-90,scale=0.65]{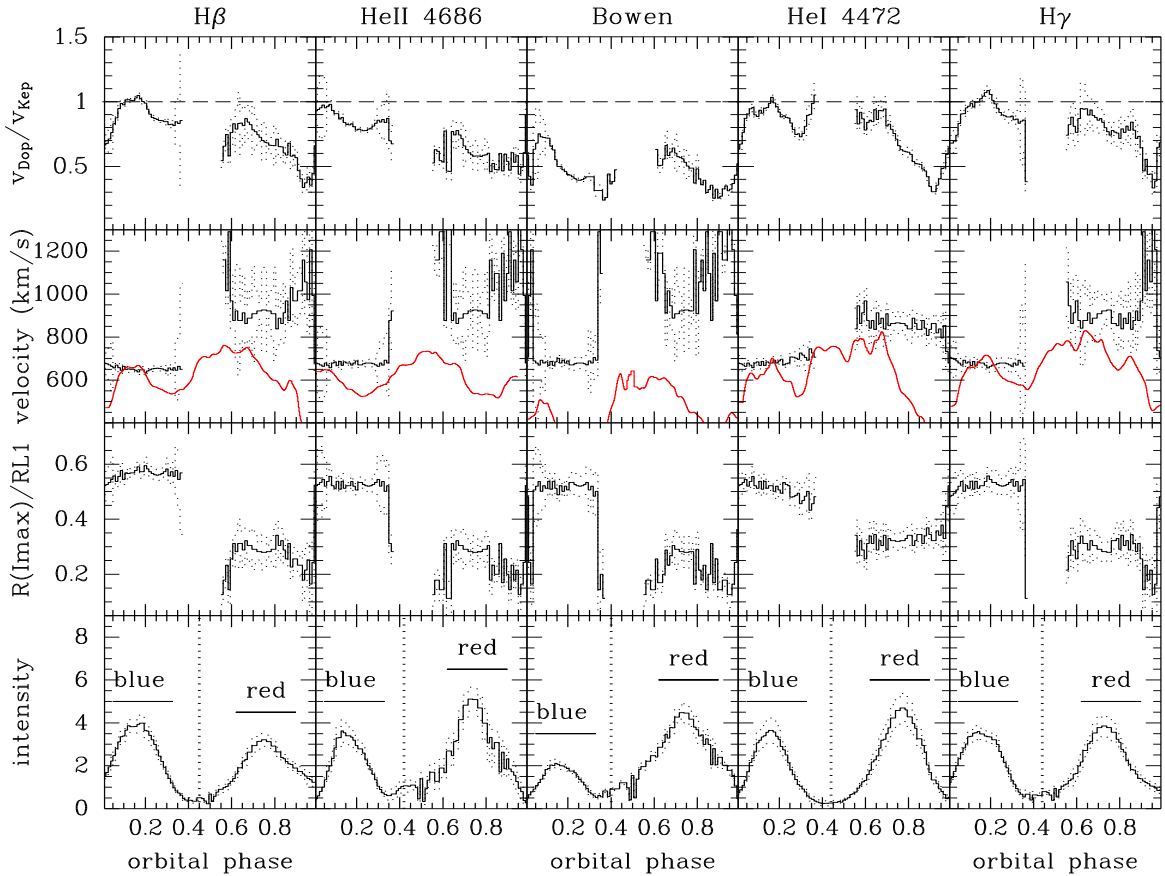}
\caption{ The dependency with binary phase of the maximum intensity 
   (lowermost panels), radius and corresponding Keplerian velocity at 
   maximum intensity (second and third rows of panels from the bottom) 
   as derived from the line eclipse maps.  
   Dotted lines depict the $\pm 1 \sigma$ limits in these distributions.  
   The velocities were computed assuming $M_1= 1.02\; M_\odot$ and 
   $R_{L1}= 0.81\; R_\odot$ (Marsh \& Horne 1990), and the intensities 
   are plotted in an arbitrary scale.  The orbital dependency of the 
   Doppler velocity of maximum intensity (extracted from the Doppler 
   maps of Fig.~\ref{fig3}) is shown as a solid grey line in the velocity 
   panels.  The location of the spiral arms are indicated in the 
   lowermost panels by horizontal bars labeled ``blue'' (outer spiral) 
   and ``red'' (inner spiral).  Vertical dotted lines in the lowermost 
   panels mark the phase of minimum intensity of the valley in between 
   the spirals and are used to estimate the spirals opening angle (see 
   section~\ref{discuss}).  The uppermost panels show the ratio of the 
   observed velocity (Doppler map) to the Keplerian velocity at the same 
   radius (eclipse map).  A horizontal dashed line in each panel depicts 
   the reference unity velocity ratio. }
\label{fig4}
\end{figure*}

Fig.~\ref{fig4} shows the azimuthal intensity distribution $I_{max}$, 
radius $R(I_{max})$ and corresponding Keplerian velocity at maximum 
intensity $v_{kep}[R(I_{max})]$.  Azimuths are expressed in terms of 
orbital phase.  These are measured from the line joining both stars 
and increase clockwise for the eclipse maps of Fig.\ref{fig3}.  
The uncertainties in $I_{max}$, $R(I_{max})$ and $v_{kep}[R(I_{max})]$ 
are derived from the Monte Carlo simulations (see section \ref{cluz}).  
Dotted lines in Fig.\ref{fig4} depict the 1-$\sigma$ limits in these 
distributions.
%

The two-armed asymmetric structures in the eclipse maps lead to a 
double-humped shape in the azimuthal intensity distribution.  The valleys 
in the azimuthal intensity distribution trace the region in between the 
spirals, where $R(I_{max})$ is meaningless.  Accordingly, the 
distributions of $R(I_{max})$ and $v_{kep}[R(I_{max})]$ are cut-off in 
the phase range where the intensity drops below 10 per cent of the peak 
intensity.  The maximum intensity along the outer, blue arm occurs at
a radius of $0.55\pm 0.05 \;R_{L1} \; (v_{kep}= 670\pm 40\; km\,s^{-1})$, 
whereas that of the inner, red arm is at $0.30\pm 0.10 \;R_{L1} \;
(v_{kep}= 900\pm 100\;km\,s^{-1})$.  The position (and velocity) of
the blue arm is more precisely measured than that of the red arm.
The location, radial and azimuthal range of the spiral arms are the 
same for all lines -- from the low-excitation Balmer lines to the high 
excitation He\,II and Bowen blend lines -- within the uncertainties.

A similar procedure was applied to the Doppler maps of Fig.~\ref{fig3} 
to compute the observed velocity at the maximum intensity as a function 
of orbital phase.  The results are plotted as solid grey lines in the 
velocity panels of Fig.~\ref{fig4}.  The results for the Bowen blend 
line are less reliable than for the other lines as the corresponding
Doppler map is quite noisy (Fig.~\ref{fig3}).  We shall regard the
results of the following analysis for this line as merely illustrative.

The upper panels of Fig.~\ref{fig4} show the ratio of the observed 
velocity (Doppler map) to the Keplerian velocity at the same radius 
(eclipse map).  A horizontal dashed line in each panel depicts the 
reference unity velocity ratio.  This comparison indicates that the 
velocity of the emitting gas along the spiral arms is generally lower 
than the Keplerian velocity at that radius.  The difference is small for 
the blue arm, which shows velocities 10-15 per cent lower than Keplerian.  
However, the discrepancy is large for the red arm, with differences 
reaching up to 40 per cent.  This indicates that the velocity field 
along the spiral arms is not Keplerian, in agreement with the results 
of BHS.  This result seems in line with the interpretation of these 
asymmetric structures as caused by tidally-induced spiral shocks, 
because the heated (and emitting) post-shock gas is expected to move
at velocities lower than the Keplerian velocity of the non-shocked 
disc orbiting gas at the same radius.

Nevertheless, hydrodynamical simulations by Steeghs \& Stehle (1999)
suggest that the departures from a Keplerian flow along the spiral 
pattern increase with radius, and that one should expect the velocity 
discrepancy to be larger for the outer, blue arm -- just the opposite 
of what is observed.  The larger velocity discrepancy (and possibly
also the smaller distance with respect to disc centre) of the red 
arm may be a consequence of its interaction with the infalling
gas stream.  In this regard, the hydrodynamical simulations by
Bisikalo et~al. (1998) and Makita et~al. (2000) show that the gas 
stream may penetrate the disc, forming a bow shock that interacts and 
mixes with the red arm to enhance emission along a spiral 
pattern close to the disc centre.

\section{Discussion} \label{discuss}

\subsection{Measuring the opening angle of the spirals} 
\label{measure}

The shape of the azimuthal intensity distribution and the position of 
maxima/minima intensities are defined by the opening angle of the 
spiral arms (see the Appendix).  For a fixed orientation of the spiral 
arms (as expected for tidally-induced spirals), the azimuth 
(binary phase) of maximum/minimum intensity increases as the opening
angle of the spiral decreases.  We performed simulations with artificial
brightness distributions containing spiral arms to calibrate this
relation.  These are described in the Appendix.  It is therefore possible
to infer the opening angle of the spiral arms from the measured binary
phase of minimum in the intensity distribution of Fig.~\ref{fig4}.

We fitted a parabola to the lower portion of the $I_{max}$ distribution 
to find an average orbital phase of minimum of $0.43\pm 0.02$ cycles.  
From this measurement, we estimate the opening angle of the spiral arms 
in our data to be $\phi= 25^o\pm 3^o$.
We performed similar analysis with the maximum intensity distributions
of BHS and BHT to find opening angles of $\phi= 34^o\pm 3^o$ and
$\phi= 14^o\pm 4^o$, respectively.  Our data were collected 5-6 days 
after the onset of the 1994 October outburst.  Those of BHS and BHT
were obtained, respectively, 3 and 9 days after the start of the 
corresponding outbursts.  

\begin{figure}
\includegraphics[bb=0.5cm 1cm 19.5cm 18cm,angle=-90,scale=0.36]{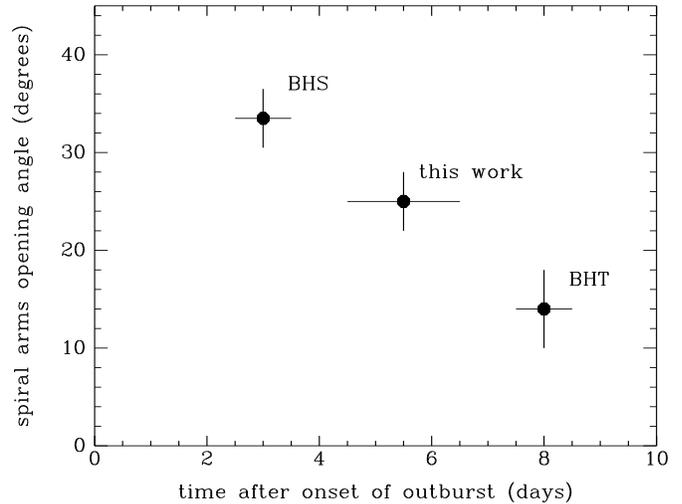}
\caption{ The measured opening angle of the spiral arms in IP~Peg as a 
    function of outburst stage. }
\label{fig5}
\end{figure}

Fig.~\ref{fig5} shows the evolution of the measured opening angle of 
the spiral arms in IP~Peg with the time from the start of the outburst.
%
%
There is a clear correlation between the measured spiral opening angle 
and the stage of the outburst, in the sense that the later the outburst 
stage (fainter, cooler and smaller accretion disc), the more tightly 
wound the spiral arms are.  This is supported by 2-D and 3-D 
hydrodynamical simulations of accretion discs, which indicate that the 
opening angle of the spiral arms is correlated to the disc temperature, 
with cooler discs showing smaller opening angles (e.g., Makita et~al. 
2000; Sato et~al. 2003).

\subsection {Spiral structures: shocks or irradiation effects?}

The spiral structures seen in IP~Peg and a few other outbursting dwarf 
novae have usually being interpreted as an observational confirmation 
of tidally induced spiral shocks in accretion discs (Steeghs 2001 and 
references therein).  Smak (2001) and Ogilvie (2002) proposed that these
spiral structures are not evidence of spiral shocks, but are caused by 
irradiation of tidally thickened sectors of the outer disc by the hot, 
inner disc regions.  

Smak's (2001) model predicts that the blue arm is systematically brighter
than the red arm.  There is no observational support for this claim.
Our analysis of 5 different emission lines show that the ratio of the
brightness of the two arms is line-dependent and probably also time
dependent. For example, both arms have comparable brightness in the
He\,II $\lambda 4686$ Doppler tomogram of the first night of observations,
whereas the blue arm appears stronger than the red arm in the Doppler
tomogram of the same line in the second night of observations (see 
Fig.\,4 of Morales-Rueda et~al. 2000).  Our data show no clear evidence 
that one arm is systematically brighter than the other one.

Our results may be used to set further constraints on both 
interpretations.

First, the eclipse mapping procedure projects any emission produced at 
height $z$ above the disc back along the inclined line of sight to the 
point where it pierces the disc plane, at a distance $z \tan i$ away 
from the secondary star with respect to its true position (Baptista 
et~al. 1995).  Conversely, if the spiral structures are produced in 
vertically-extended disc regions, their true positions would be at a 
distance $z \tan i$ towards the secondary star.
For the high inclination of IP~Peg, even a small vertical height 
$z=0.03 \,R_{L1} \simeq 1.7 \times 10^9\;cm$ moves the spiral arms by
$\Delta x \simeq 0.2 \;R_{L1}$ towards the secondary star with respect 
to their positions in the eclipse maps.  This displacement is enough 
to shift both arms to the front side of the disc (the disc hemisphere 
closest to the secondary star).
However, the effects raised by the tides from the secondary star are 
point-symmetric with respect to the white dwarf, for both the shocks 
and the irradiation interpretations (Steeghs \& Stehle 1999; Makita 
et~al. 2000; Smak 2001; Ogilvie 2002).  It would be hard to explain 
the resulting highly asymmetric geometrical configuration within the 
framework of tides induced by the secondary star.  

In order to avoid completely braking the point-symmetry of the
arms with respect to the white dwarf, they must be at heights lower 
than $z \simeq 0.015 \;R_{L1} \simeq 8\times 10^8\;cm$.
But this is of the order of (or even lower than) the expected thickness
of the accretion disc at the radial position of the arms 
\footnote{For a disc opening angle $\alpha \simeq 2^o-3^o$, the disc half
thickness $H$ at a distance $R\simeq 0.4\;R_{L1}$ from its centre will be 
$H= R \tan\alpha \simeq (0.014-0.02)\;R_{L1} \simeq (0.8-1.1) \times 
10^9\;cm$.}.
Thus, our results suggest that the site of the line emission along the 
spirals cannot be too far above the orbital plane.  There is no evidence 
of a measurable disc thickening effect.

Second, a consequence of the irradiation model is that mainly the side 
of the thickened disc regions facing the hot inner disc would be 
irradiated.  When the thickened disc structure is viewed from a highly
inclined line of sight towards the secondary star (i.e., around eclipse) 
the illuminated red arm would be seen mostly face-on, but the blue arm 
would suffer self-occultation (or would be seen with a much lower 
effective area).  Thus, in the irradiation model one expects that 
the red arm appears systematically (and possibly significantly) 
brighter than the blue arm around eclipse (and this brightness 
difference would be transported to any eclipse map).  Eclipse mapping
results offer no support for this prediction.

Finally, while sub-Keplerian velocities along the spiral pattern are 
expected in the spiral shock model, there is no immediate explanation 
for them in the irradiation model.

Therefore, the combined results from the Doppler tomograms and eclipse
maps seem to favour the spiral shock interpretation of the observed 
asymmetric structures.

\section {Conclusions} \label{conclusions}

We reanalyzed time-resolved spectroscopy of IP~Pegasi in outburst
with eclipse mapping techniques in order to investigate the location
and geometry of the observed spiral structures. We were able to obtain
an improved view of the spiral arms with the aid of light
curves extracted in velocity bins matching the observed range of
velocities of the spiral arms and a double default map tailored
for reconstruction of asymmetric structures.

Two-armed spiral structures are clearly seen in all eclipse maps.
The arms are located at different distances from the disc centre.
The ``blue'' arm is farther out in the disc ($R= 0.55 \pm 0.05\;R_{L1}$) 
than the ``red'' arm ($R= 0.30 \pm 0.05\;R_{L1}$).  The observed 
difference in radius is significant at the 3-$\sigma$ level.
The ``red'' arm is stronger in the Bowen blend, whereas the ``blue'' 
arm is stronger in $H\beta$.  The two arms have comparable brightnesses 
in the other lines.  The brightness ratios derived from the eclipse maps 
and Doppler tomograms are consistent with each other.

There is evidence that the velocity of the emitting gas along the
spiral pattern is lower than the Keplerian velocity for the same disc 
radius.  The discrepancy is lower in the ``blue'', outer arm -- with
the measured Doppler velocities being typically 10-15 per cent lower
than Keplerian -- but is more significant in the ``red'', inner arm
-- where the observed velocities can be 40 per cent lower than the
Keplerian velocities.

It is possible to measure the opening angle $\phi$ of the spirals from 
the azimuthal intensity distribution of the eclipse maps.  For the data
of this paper we measure $\phi= 25^o\pm 3^o$.  We performed similar
analysis with the data of BHS and BHT to find opening angles of
$\phi=34^o\pm 3^o$ and $\phi=14^o\pm 4^o$, respectively.
The difference between the measured opening angle of the spirals in
these three occasions is statistically significant at the 3-$\sigma$ 
confidence level.  
There is clear evidence that the opening angle of the spiral arms in
IP~Peg decreases towards the later stages of the outburst while the
accretion disc cools and shrinks.  This fits nicely into the expected 
evolution of a tidally driven spiral wave and it is supported by 
hydrodynamical simulations of accretion discs.

The sub-keplerian velocities along the spiral pattern and the clear
correlation between the opening angle of the spirals and the outburst
stage favors the interpretation of these asymmetric structures as 
tidally-induced spiral shocks.

It would be crucial to confirm these results with further time-lapse,
combined Doppler tomography and eclipse mapping of emission lines of 
IP~Peg along the same outburst cycle.

\begin{acknowledgements}

We dedicate this paper to the memory of our colleague and dear friend 
Emilios Harlaftis.
The INT Telescope is operated on the Island of La Palma by the Isaac
Newton Group in the Spanish Observatory del Roque de los Muchachos of
the Instituto de Astrofisica de Canarias. 
This work was partially supported by CNPq/Brazil through the research 
grant 62.0053/01-1 -- PADCT III/Milenio. RB acknowledges financial 
support from CNPq/Brazil through grants 300.354/96-7 and 301.442/2004-5. 
LMR was supported by a PPARC post-doctoral grant and by NWO-VIDI grant
639.042.201 to P. J. Groot during the course of this research.
TRM acknowledges the support of a PPARC Senior Research Fellowship.
DS acknowledges a Smithsonian Astrophysical Observatory Clay Fellowship.

\end{acknowledgements}

\appendix

\section {A double default eclipse mapping}

The default map defines the point-spread-function of the reconstructed
eclipse map.  The original default map of Horne (1985) blurs any 
asymmetric point-like source into a ring at the same disc radius and 
is well suited to recover the radial profile of broad and fairly
symmetrical disc brightness distributions.  

The limited azimuthal smearing default (Rutten et~al. 1992; Baptista 
et~al. 1996) added the capability to recover azimuthal information 
about asymmetrically located light sources such as the bright spot 
at the disc rim. 
Nevertheless, it was not designed for reconstructions of highly 
asymmetric light sources such as the spiral structures seen in the
accretion disc of IP~Peg in outburst.
The azimuthal smearing default map blurs the spiral pattern into a
"butterfly"-shaped structure (Harlaftis et~al. 2004). It is possible
to use these maps to define the range of radii and azimuths covered by 
the spirals, but it is not possible to trace the spiral pattern --
i.e., to measure the spiral opening angle $\phi$.

Harlaftis et~al.\ (2004) suggests the use of a default map which 
steers the eclipse mapping solution towards a spiral shape in order 
to improve the reconstruction of spiral structures in IP~Peg.  The idea 
is to minimize the azimuthal blur effect, allowing to trace the observed
spiral arms.

We followed this idea and created a {\em spiral default map}, 
which blurs the intensity of each point source in the eclipse 
map along a spiral pattern in the disc.  The first problem with this 
approach is that the spiral default map introduces an extra, {\em ad hoc}
parameter, namely, the opening angle of the spiral pattern $\phi$. 
Since the opening angle is not known in advance (on the contrary, this
is one of the quantities we wish to measure from the eclipse map),
a plausible approach would be to make reconstructions for a set of 
values of $\phi$ and try to determine which map is the correct one.  

In order to test the ability of the eclipse mapping method in 
recovering the proper value of $\phi$ using the spiral default map, 
we created artificial maps with two-armed logarithmic spiral patterns 
for different values of $\phi$.  The spirals have a radial width of 
$0.08\; R_{L1}$ and a radial range between 0.2-0.6 $R_{L1}$ with a 
brightness cut off $\exp[-\mid(r-r_0)/dr\mid^3]$ law.  The trace of 
the spirals is defined by,
\begin{equation}
\ln r(\theta) = \ln r(\theta_0) + \tan\phi\; (\theta-\theta_0) \;\;\; ,
\label{traco}
\end{equation}
where $r$ and $\theta$ are, respectively, the radial and azimuthal 
coordinates, $[\theta_0,r(\theta_0)]$ defines a reference point along
the spiral pattern, and $\phi$ is the spiral opening angle.

We simulated the eclipse of the artificial maps using the geometry of 
IP~Peg ($i=81^o$ and $q=0.5$, Wood \& Crawford 1986) to create light 
curves with added gaussian noise and obtained reconstructions using 
spiral default maps for a set of $\phi$ values.
These simulations show that there is no correlation between the map 
with the correct $\phi_{real}$ and the map of highest entropy in a 
sequence of reconstructions with a range of values of $\phi$.
We further tried to see if one could recover the correct value 
$\phi_{real}$ by using a spiral default map with a fixed, arbitrary 
value $\phi_0$.  
In all cases the asymmetric arcs in the reconstruction are steered 
towards $\phi_0$ instead of towards $\phi_{real}$.
The conclusion is that using a spiral default map is not useful in
investigating spiral structures in eclipse maps, unless we know in
advance the opening angle of the spirals.

With the usual default map of limited azimuthal smearing, the
ability to trace the spiral pattern is hampered by the azimuthal 
smearing effect, which is responsible for the ``butterfly'' shape 
of the reconstructed spiral arms (Harlaftis et~al.\ 2004).
For an azimuthal gaussian blur width of $\Delta\theta= 30^o$, all 
point sources in the eclipse map will be azimuthally smeared by 
$\simeq 90^o$.  An obvious way to minimize the azimuthal smearing effect
and improve the reconstruction of asymmetric structures would be to
reduce $\Delta\theta$.  This corresponds to relaxing the fundamental 
constraint that the disc brightness distribution has to be the most 
axi-symmetric as possible.  However, eclipse maps obtained with 
lower values of $\Delta\theta$ start to show criss-crossed arcs
at the position of bright compact sources.  The reason it that, without 
the axi-symmetry constraint, the intensity of a compact source is 
distributed along the arcs that outline the shadow of the secondary 
star at the corresponding ingress/egress phases (e.g., Horne 1985).  
The need to avoid this physically unreasonable mapping is what prompted
Horne (1985) to propose the default map with azimuthal smearing.

A default map which tries to {\em avoid} building structure along the
ingress/egress arcs of the secondary star would be particularly 
interesting for the reconstruction of asymmetric sources, because
in this case $\Delta\theta$ could be reduced without the above undesired 
side effect. The approach of multiple default maps of Spruit (1994) 
is very useful in this regard.

Following the prescription of Spruit (1994), the concept of the default
function $D(j)$ can be generalized to the case of multiple functions
by,
\begin{equation}
D(j) = e^{( \sum_i n_i \ln D_i )} = \prod_i \; D_i^{n_i} \;\;\; , \;\;\;
\sum_i n_i = 1 \;\;\; ,
\end{equation}
where $n_i$ is the exponent of the individual default function $D_i$.
If $n_i>0$, the method will steer the intensities $I(j)$ towards the
default map $D_i$, whereas if $n_i<0$ the method with steer the intensities
away from the default map $D_i$.

Here we implement a default function consisting of two default maps: 
a standard default map of limited azimuthal smearing $D_{az}$ with a 
positive index $n$, and a default map of criss-crossed arcs $D_{criss}$
with a {\em negative} exponent ($1-n$),
\begin{equation}
D(j) = [D_{az}(j)]^n \times [D_{criss}(j)]^{(1-n)} \;\;\; , \;\;\; n>1
\;\;\; .
\end{equation}
This is the equivalent of simultaneously requesting for the most 
axi-symmetric solution with the least amount of structures along the
ingress/egress arcs of the secondary star.
The default map of criss-crossed arcs is defined by,
\begin{equation}
D_{criss}(j) = \frac{\sum_k w_{jk} I_k}{\sum_k w_{jk}} \;\;\; ,
\end{equation}
where
\begin{equation}
w_{jk} = \exp\left[ - \frac{1}{2} \left( 
\frac{ \phi_i(j) - \phi_i(k) }{\Delta\phi} \right)^2 \right]
+ \exp\left[ -\frac{1}{2} \left( 
\frac{ \phi_e(j) - \phi_e(k) }{\Delta\phi} \right)^2 \right] ,
\end{equation}
$I_k$ is the intensity of pixel $k$, $\phi_i(j)$ and $\phi_e(j)$ are 
the ingress and egress phases of pixel $j$ respectively, and 
$\Delta\phi$ is the binary phase blur width.
Simulations show that good results are obtained for $n \simeq 1.1-1.2$ 
and $\Delta\phi \simeq 0.002 - 0.004$ cycles. 

\begin{figure}
\includegraphics[bb=1cm 1.5cm 18cm 25cm,scale=0.53]{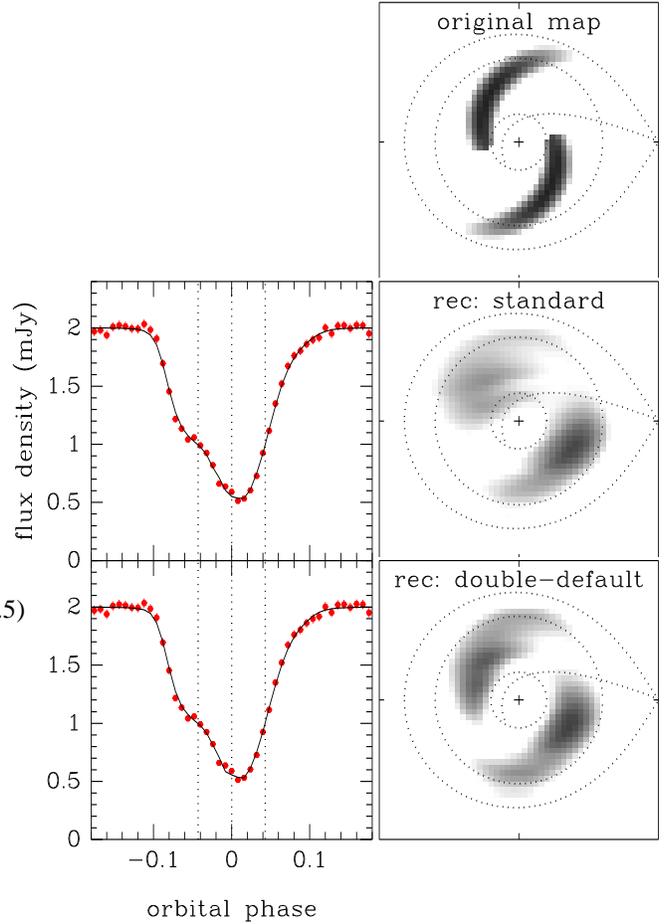}
\caption{ Comparison of reconstructions obtained with the standard 
   eclipse mapping (single default map) and the double-default eclipse 
   mapping.  The top panel shows the original map with two spiral arms of
   opening angle $\phi=30^o$ in a logarithmic greyscale.  The left panels 
   depict the simulated data (dots with error bars) and model light 
   curves (solid lines) in each case.  Vertical dotted lines indicate the 
   phases of mid-eclipse and of ingress/egress of the disc centre.
   The right panels show the reconstructions obtained with the standard 
   eclipse mapping (middle panel) and with the double default mapping 
   (lower panel) in the same logarithmic greyscale. The notation is similar
   to that of Fig.\ref{fig3}.}
\label{apfig1}
\end{figure}

This double-default function is tailored for reconstructions of asymmetric
brightness distributions.  It yields no perceptible improvement in 
the quality of reconstructions of broad, fairly symmetric brightness
distributions found in usual eclipse mapping applications.
However, it does make a difference for reconstructions of spiral arms.
Fig.~\ref{apfig1} compares a reconstruction of a brightness distribution
with two spiral arms as obtained with the standard single default map
of limited azimuthal smearing with that obtained with the double default
map.
%
%
The two reconstructions were obtained with $\Delta\theta= 10^o$.
Because of the reduced $\Delta\theta$ value, a criss-crossed structure 
develops in the reconstruction with the single default map. It can be
seen as a bright arc running across the upper left spiral arm. This
arc outlines the shadow of the secondary star at mid-ingress phase of
that spiral arm.  This artifact is largely suppressed in the 
reconstruction with the double default map.  Furthermore, the 
``butterfly'' shape of the reconstruction with the single default map 
is replaced by a better defined trace of the spirals in the 
reconstruction with the double default map.

\begin{figure*}
\includegraphics[bb=1cm 2cm 16cm 24cm,scale=0.75]{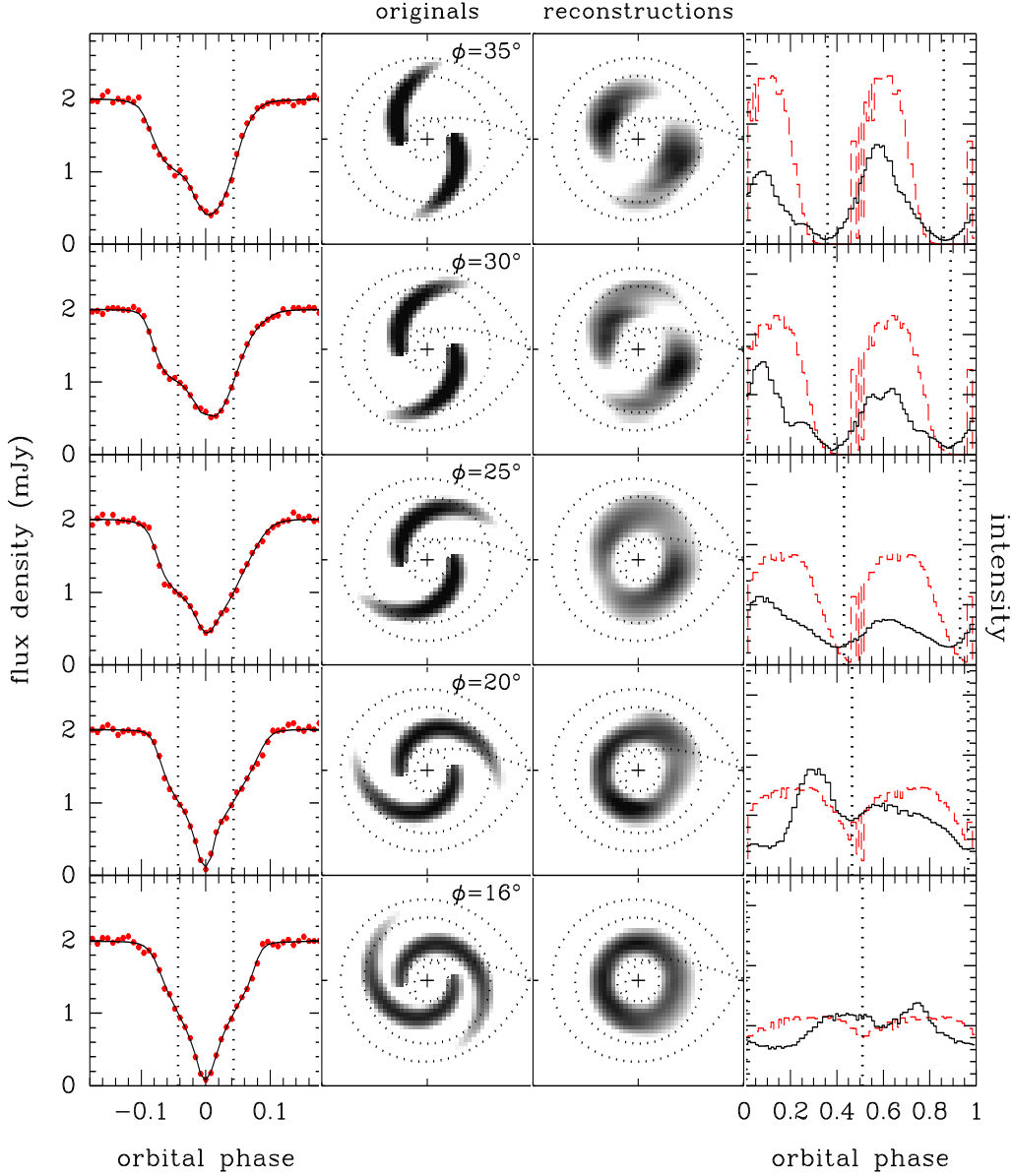}
\caption{ Sequence of brightness distributions with spiral arms of
    different opening angles. Left panel: simulated noisy light curves 
    (S/N=50, gray dots with error bars) and eclipse mapping model light
    curves (solid lines).  Vertical dotted lines mark the ingress and 
    egress phases of the disc centre.  Middle panels: original maps (left)
    and eclipse mapping reconstructions with the double-default map 
    (right). The reconstructions were obtained with $\Delta\theta= 10^o$. 
    The notation is similar to that of Fig.~\ref{apfig1}.  Right panels: 
    the azimuthal dependency of the maximum intensity of the original
    maps (dashed line) and of the reconstructions (solid line). Vertical
    dotted lines mark the azimuths of minimum light in each case. }
\label{apfig2}
\end{figure*}

Fig.~\ref{apfig2} shows the performance of the double-default eclipse
mapping for the reconstruction of two-armed spiral brightness distributions
at different opening angles.  The left hand panels display the data and
model light curves. It is seen that the shape of the eclipse changes 
significantly as the opening angle of the spirals is reduced.
Original and reconstructed maps are shown in the middle two columns.
The right-hand panels show the distributions of the maximum intensity
in the map as a function of binary phase (or azimuth, see BHS). 
The distribution of the original map is shown as a dashed gray line 
while that of the reconstruction is indicated by a solid line. Vertical 
dotted lines mark the azimuths of minimum intensity in each case.
Orbital phases are measured with respect to the inner lagrangian point
L1 and increase clockwise.
%
%

Wide open spirals ($\phi > 25^o$) produce narrow, well defined maxima
and minima in the azimuthal intensity distribution.  The corresponding
eclipse maps show well defined and separated asymmetric arcs with a 
reasonably good reconstruction of the shape, radial and azimuthal range 
of the spirals.  As the opening angle is reduced, the maxima of the 
azimuthal intensity distribution become broader while the minima become
narrower.  Because of the winding up of the spiral arms, both maxima 
and minima intensities move towards higher orbital phases in a monotonic,
fairly linear way.  The shape of the azimuthal distribution and the 
position of maxima/minima intensities are defined by the opening angle 
of the spiral arms.  Thus, it is possible to estimate the opening angle 
of the spirals by measuring the position of maxima/minima intensity in 
the azimuthal distribution.

The maxima of the azimuthal intensity distribution are affected by 
noise in the light curve and by residual criss-crossed arc effect and 
are less reliably recovered in the reconstructions.
On the other hand, the minima of the azimuthal intensity distribution 
trace the region in between the spirals.  Because there is essentially 
no flux at these azimuths, their positions are less prone to be distorted 
by noise.  Also, the minima become narrower as the opening angle 
decreases and are, therefore, easier to measure in the limit of low 
$\phi$'s.  We therefore choose to measure the opening angle of the 
spirals from the azimuths of minimum intensity.

\begin{figure}
\includegraphics[bb=1cm 1.5cm 18cm 24cm,scale=0.53]{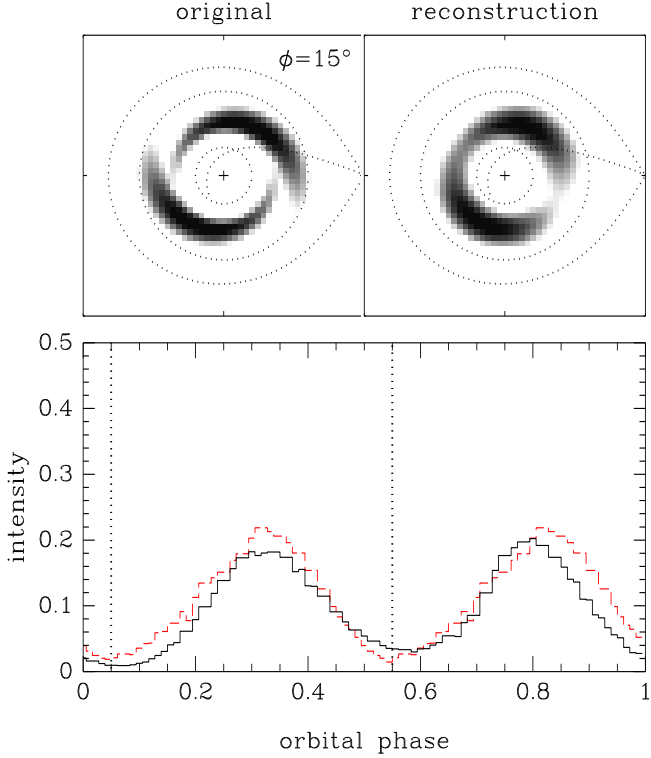}
\caption{ Reconstruction of a brightness distribution with spiral arms of
    $\phi=15^o$ and radial range $0.3-0.55\;R_{L1}$.  Top: original (left)
    and reconstructed (right) maps. Bottom: the azimuthal dependency of 
    the maximum intensity of the original map (dashed line) and of the 
    reconstruction (solid line).  The notation is the same as in 
    Fig.~\ref{apfig2}. }
\label{apfig3}
\end{figure}

For small opening angles the two spiral arms become so tightly wound 
that they start to overlap in azimuth.  Because of the (reduced, but still 
present) azimuthal smearing effect, they become indistinguishable in the 
reconstructions and it is no longer possible to use the azimuthal 
intensity distribution to estimate the opening angle of the spirals.
For the simulations presented in Fig.~\ref{apfig2}, this occurs at
$\phi_{min}\simeq 16^o$.
We shall emphasize that the minimum opening angle that is possible 
to measure depend on the radial extension of the spiral arms.
If the spiral pattern does not extend as far inside the disc as those
in Fig.~\ref{apfig2} it is possible to measure $\phi$ for spirals with 
smaller opening angles than the limit of the above example. 
This is illustrated in Fig.~\ref{apfig3}, which shows the results for
a two-armed spiral pattern distribution with $\phi=15^o$ and a reduced 
radial range of $0.3-0.55 \;R_{L1}$.  The azimuthal intensity distribution
is easily recovered in this case, allowing a precise measurement of the
azimuth of minimum intensity.  The reconstructed brightness distribution 
is similar to those of the IP~Peg eclipse maps of BHT.
%
%

From our simulations we find the following empirical relation between
the spiral opening angle $\phi$ and the observed orbital phase of minimum
intensity $\Phi_{orb}(I_{min})$,
\begin{equation}
\phi(\hbox{degrees}) = \frac{23.25}{\Phi_{orb}(I_{min})} - 29.6 \;\; .
\end{equation}

In summary, by using a double default map and by measuring the minima
of the azimuthal intensity distribution in the resulting eclipse map it 
is possible to estimate the opening angle $\phi$ of the spiral arms in 
IP~Peg.


\begin{thebibliography}{}

\bibitem [1994]{bap94} Baptista, R., et al. 1994, ASP Conference Series 56, 
    ed. A. Shafter, ASP: San Francisco, p. 259
\bibitem [2000]{bhs00} Baptista, R., Harlaftis, E. T., \& Steeghs, D. 2000,
    \mnras, 314, 727
\bibitem [2002]{bht02} Baptista, R., Haswell, C. A., \& Thomas, G. 2002,
    \mnras, 334, 198
\bibitem [1995]{bap95} Baptista, R., Horne, K., Hilditch, R., Mason, K. O., 
    \& Drew, J. E. 1995, ApJ, 448, 395
\bibitem [1993]{bst93} Baptista, R., \& Steiner, J. E. 1993, A\&A, 277, 331
\bibitem [1996]{bsh96} Baptista, R., Steiner, J. E., \& Horne, K. 1996, 
    \mnras, 282, 99
\bibitem [1998]{bisi} Bisikalo, D. V., Boyarchuk, A. A., Chechetkin, V. M.,
    Kuznetsov, O. A., \& Molteni, D. 1998, \mnras, 300, 39
\bibitem [1998]{godon} Godon, P., Livio, M., \& Lubow, S. 1998, \mnras, 
    295, L11
\bibitem [1999]{h99} Harlaftis, E.~T. 1999, \aap, 346, L73
\bibitem [2004]{har04} Harlaftis, E.~T., Baptista, R., Morales-Rueda, L., 
    Marsh, T. R., \& Steeghs, D. 2004, \aap, 417, 1063
\bibitem [1999]{har99} Harlaftis, E.~T., Steeghs, D., Horne K., Mart\'{\i}n
    E., \& Magazz\'u A. 1999, \mnras, 306, 348
\bibitem [1985]{hrn85} Horne, K. 1985, \mnras, 213, 129
\bibitem [2000]{makita} Makita, M., Miyawaki, K., \& Matsuda, T. 2000, 
    \mnras, 316, 906
\bibitem [1988]{msr88} Marsh, T. R., \& Horne, K. 1988, \mnras, 235, 269
\bibitem {34} Marsh, T. R., \& Horne, K. 1990, \apj, 349, 593
\bibitem [2000]{mor00} Morales-Rueda, L., Marsh, T. R., \& Billington, I. 
    2000, \mnras, 313, 454
\bibitem [2002]{og} Ogilvie, G. I. 2002. \mnras, 330, 937 
\bibitem [1992]{rut92} Rutten, R. G. M., van Paradijs, J., \& Tinbergen, J.
    1992, \aap, 254, 159
\bibitem [2005]{saito} Saito, R., Baptista, R., \& Horne, K. 2005. \aap, 
    433, 1085
\bibitem [2003]{sat03} Sato, J., Sawada, K., \& Ohnishi, N. 2003, \mnras, 
    342, 593
\bibitem {saw86} Sawada, K., Matsuda, T., \& Hachisu, I. 1986, \mnras, 
    219, 75 
\bibitem [2001]{smak} Smak, J. I. 2001, Acta Astron., 51, 295
\bibitem {spruit94} Spruit, H. C. 1994, \aap, 289, 441
\bibitem {steeghs01} Steeghs, D. 2001, Astrotomography, Indirect Imaging
    Methods in Observational Astronomy, Lecture Notes in Physics, vol.
    573, p. 45, Springer-Verlag
\bibitem [1997]{shh97} Steeghs, D., Harlaftis, E. T., \& Horne, K. 1997, 
    \mnras, 290, L28
\bibitem [1996]{steeghs96} Steeghs, D., Horne, K., Marsh, T. R., \& Donati,
    J. F. 1996, \mnras, 281, 626
\bibitem [1999]{ss99} Steeghs, D., \& Stehle, R. 1999, \mnras, 307, 99
\bibitem [1993]{wol93} Wolf, S., Mantel, K. H., Horne, K., Barwig, H., 
    Shoembs, R., \& Baernbantner, O. 1993, \aap, 273, 160
\bibitem [1986]{wcr86} Wood, J. E., \& Crawford, C. S. 1986, \mnras, 222, 645

\end{thebibliography}
\end{document}